\documentclass[aps,showpacs,showkeys,amsmath,amssymb,pra,reprint,superscriptaddress,lengthcheck]{revtex4-1}

\usepackage{dcolumn}
\usepackage{amsmath}
\usepackage{amssymb}
\usepackage{color}
\usepackage{graphicx}
\usepackage{bm}
\begin{document}

\title{Rayleigh scattering of twisted light by hydrogenlike ions}

\author{A.~A.~Peshkov}
\email{anton.peshkov@uni-jena.de}
\affiliation{Helmholtz-Institut Jena, D-07743 Jena, Germany}

\author{A.~V.~Volotka}
\affiliation{Helmholtz-Institut Jena, D-07743 Jena, Germany}
\affiliation{Department of Physics, St.~Petersburg State University, 198504 St.~Petersburg, Russia}

\author{A.~Surzhykov}
\affiliation{Physikalisch-Technische Bundesanstalt, D-38116 Braunschweig, Germany}
\affiliation{Technische Universit\"at Braunschweig, D-38106 Braunschweig, Germany}

\author{S.~Fritzsche}
\affiliation{Helmholtz-Institut Jena, D-07743 Jena, Germany}
\affiliation{Theoretisch-Physikalisches Institut $\&$ Abbe Center of Photonics, Friedrich-Schiller-Universit\"at Jena, D-07743 Jena, Germany}

\keywords{elastic scattering, atoms, Dirac equation, transitions, polarization, optical vortex, OAM}
\pacs{31.10.+z, 32.80.Wr, 42.50.Tx}

\date{\today}

\begin{abstract}
The elastic Rayleigh scattering of twisted light and, in particular, the polarization (transfer) of the scattered photons have been analyzed within the framework of second-order perturbation theory and Dirac's relativistic equation. Special attention was paid hereby to the scattering on three different atomic targets: single atoms, a mesoscopic (small) target, and a macroscopic (large) target, which are all centered with regard to the beam axis. Detailed calculations of the polarization Stokes parameters were performed for C$^{5+}$ ions and for twisted Bessel beams. It is shown that the polarization of scattered photons is sensitive to the size of an atomic target and to the helicity, the opening angle, and the projection of the total angular momentum of the incident Bessel beam. These computations indicate more that the Stokes parameters of the (Rayleigh) scattered twisted light may significantly differ from their behaviour for an incident plane-wave radiation.
\end{abstract}

\maketitle

\section{\label{sec:intro}Introduction}

The elastic scattering of photons at the bound electrons of atoms or ions, commonly known as Rayleigh scattering, has been intensively explored over the past decades \cite{Brini/NC:1959, Roy/PRA:1986, Smend/PRA:1987}. From a theoretical viewpoint, the Rayleigh scattering has attracted much interest as one of the simplest second-order quantum electrodynamical (QED) process \cite{Kane/PP:1986}. From a practical viewpoint, detailed knowledge of the properties of elastically scattered photons has been found important for applications in material research \cite{Sfeir/S:2004}, medical imaging \cite{Elshemey/PMB:1999}, and astrophysics \cite{Maeda/AJ:2012}. 

In the past, a large number of experimental and theoretical studies have been performed in order to understand how the electronic structure of atoms affects the polarization of the Rayleigh-scattered photons \cite{Somayajulu/JPA:1968, Chitwattanagorn/JPG:1980, Kissel/PRA:1980, Manakov/JPB:2000, Safari/PRA:2012, Surzhykov/PRA:2013, Surzhykov/JPB:2015, Lin/PRA:1975, Volotka/PRA:2016}. In particular, the linear polarization of the elastically scattered light has been measured directly by Blumenhagen \textit{et al.}\ at the PETRA III synchrotron at DESY \cite{Blumenhagen/NJP:2016}. This experiment was performed for a gold target with a highly linearly polarized incident plane-wave radiation. Until the present, however, very little is known about the Rayleigh scattering of twisted (or vortex) light beams. When compared to plane-wave radiation, such twisted photons have a helical wavefront and carry a well-defined projection of the orbital angular momentum (OAM) upon their propagation direction \cite{Harris/Nature:2015, Bliokh/PR:2015}. In addition, the transverse intensity profile of the twisted beams exhibits a ringlike pattern with a dark spot (vortex) at the center \cite{Andrews:2013}. In experiments, twisted (Bessel) beams can nowadays be readily produced by means of spatial light modulators \cite{Walde/OC:2017} or axicons \cite{Dota/PRA:2012, Brzobohaty/OE:2008, Choporova/PRA:2017}. During recent years a number of studies have shown that the OAM and the intensity profile of these twisted beams may affect different fundamental light-matter interaction processes such as the Compton scattering \cite{Jentschura/PRL:2011, Stock/PRA:2015, Sherwin/PRA:2017}, photoexcitation \cite{Surzhykov/PRA:2015, Schmiegelow/NC:2016, Afanasev/PRA:2013, Peshkov/PS:2016, Quinteiro/PRA:2017, Peshkov/PRA:2017} and photoionization \cite{Matula/JPB:2013, Surzhykov/PRA:2016} of atoms, the generation of electric currents in quantum rings \cite{Quinteiro/OE:2009} and molecules \cite{Koksal/CTC:2017}, electromagnetically induced transparency \cite{Radwell/PRL:2015}, four-wave mixing in atomic vapors \cite{Akulshin/OL:2015}. One might therefore expect that the``twistedness''of incoming radiation will affect also the polarization of outgoing photons in the Rayleigh scattering.

In the present work, we analyze theoretically the behavior of the polarization Stokes parameters of scattered photons for the elastic scattering of twisted Bessel light. Here we restrict ourselves to the nonresonant Rayleigh scattering of light by hydrogenlike ions in their ground state, and especially by C$^{5+}$ ions. In Sec. \ref{sec:theory}, we shall consider and derive the Stokes parameters within the framework of second-order perturbation theory and the density matrix approach. Three different ``experimental'' scenarios are considered here for the scattering of the incident Bessel beam at: (i) a single atom, (ii) a mesoscopic (small) or (iii) a macroscopic (large) atomic target, and which are all assumed to be centered on the beam axis. Results of our calculations for the Bessel beams with different polarizations, opening angles, and projections of the total angular momentum (TAM) are presented in Sec. \ref{sec:results} and are compared with those for incident plane-wave radiation. These results demonstrate that the scattering of twisted light may lead to well detectable changes in the polarization of scattered photons. Finally, a summary and outlook are given in Sec. \ref{sec:summary}.

Atomic units $( \hbar \,= 4\pi\varepsilon_0 \,=\, e \,=\, m_{e}=1, \, c=1/\alpha )$ are used throughout the paper unless stated otherwise.

\section{\label{sec:theory}Theory}
\subsection{\label{sec:bessel}Vector potential of Bessel light beams}

Before we consider the Rayleigh scattering of twisted Bessel beams on atoms, let us first define and explain such beams of light. In general, all the properties of light can be described by means of the vector potential. For a Bessel beam with a well-defined helicity $\lambda_1$, longitudinal momentum $k_{z_1}$, (modulus of the) transverse momentum $\varkappa$, photon energy $\omega = ck_1 = c\sqrt{k_{z_1}^2 + \varkappa^2}$, as well as the projection $m$ of the total angular momentum (TAM) upon its propagation ($z$) direction, for instance, the vector potential is given by \cite{Surzhykov/PRA:2016}
\begin{align}
\label{eq:vec1}
	\bm{A}^{\text{tw}} (\bm{r}) = \int a_{\varkappa m} (\bm{k}_{\perp_1}) \, \bm{e}_{\bm{k}_1 \lambda_1} \, e^{i \bm{k}_1 \bm{r}} \, \frac{d^2 \bm{k}_{\perp_1}}{(2\pi)^2} \, ,
\end{align}
where the amplitude $a_{\varkappa m} (\bm{k}_{\perp_1})$ is of the form
\begin{align}
\label{eq:amp}
	a_{\varkappa m} (\bm{k}_{\perp_1}) = (-i)^{m} \, e^{i m \phi_{k_1}} \, \sqrt{\frac{2 \pi}{k_{\perp_1}}} \,  \delta (k_{\perp_1} - \varkappa) \, .
\end{align}
As seen from these expressions, such a Bessel beam can be considered also as a superposition of circularly polarized plane waves $\bm{e}_{\bm{k}_1 \lambda_1} \, e^{i \bm{k}_1 \bm{r}}$ with well-defined helicity $\lambda_1 $. Their wave vectors $\bm{k}_1$ are uniformly distributed upon the surface of a cone with an opening angle $\theta_{k_1} = \arctan (\varkappa / k_{z_1})$ and are orthogonal to the polarization vectors, $\bm{e}_{\bm{k}_1 \lambda_1} \cdot \bm{k}_1 = 0$.

Although the integral representation \eqref{eq:vec1} of the vector potential $\bm{A}^{\text{tw}} (\bm{r})$ is very convenient for atomic calculations, it is useful to perform the integration over $\bm{k}_{\perp_1}$ in Eq. \eqref{eq:vec1} explicitly, in particular for very small opening angles $\theta_{k_1} $ for which the transverse momentum is much smaller than the longitudinal one, $\varkappa \ll k_{z_1}$. Within this (so-called) paraxial approximation, this integration gives then rise to a vector potential (up to a multiplicative constant) in the form \cite{Matula/JPB:2013}
\begin{align}
\label{eq:vec2}
	\bm{A}^{\text{tw}} (\bm{r}) = \bm{\varepsilon}_{\lambda_1}  J_{m-\lambda_1} (\varkappa r_{\perp}) e^{i (m-\lambda_1) \phi}  e^{i k_{z_1} z} \, ,
\end{align}
and where $J_{m-\lambda_1} (\varkappa r_{\perp})$ denotes the Bessel function of the first kind. Substituting the polarization vector $\bm{\varepsilon}_{\lambda_1} = \bm{e}_{\bm{k}_1 \lambda_1} (\theta_{k_1} = \phi_{k_1} = 0^{\circ})$ into this expression, we see that the Bessel beam with a small opening angle has well-defined projections of the orbital $m - \lambda_1$ (OAM) and spin $\lambda_1$ (SAM) angular momenta onto the $z$ axis. However, such a decoupling of the OAM and SAM does not longer apply in the nonparaxial regime, i.e., when the opening angle $\theta_{k_1}$ becomes larger \cite{Matula/JPB:2013}.

\subsection{\label{sec:amplitude}Evaluation of the transition amplitude}
With this brief account on the vector potential of twisted Bessel beams, we can now discuss the Rayleigh scattering of such beams by hydrogenlike ions. We here begin from the Furry picture of QED, in which the electron-nucleus interaction is included into the unperturbed Hamiltonian, while the interaction with the radiation field is treated as a perturbation \cite{Kane/PP:1986}. In this picture, the properties of the scattered photons can all be obtained from the second-order transition amplitude, based on Dirac's relativistic equation. In this framework, the amplitude is given by \cite{Surzhykov/PRA:2013, Akhiezer:1965}
\begin{widetext}
\begin{align}
\label{eq:matrix1}
	\mathcal{M}_{m_f m_i}^{\lambda_2 \lambda_1} (\bm{b}) = &\sum_{n_{\nu} j_{\nu} m_{\nu}} \frac{\langle n_f j_f m_f \vert \bm{\alpha} \cdot \bm{A}^{\text{pl} *} (\bm{r}) \vert n_{\nu} j_{\nu} m_{\nu} \rangle   \langle n_{\nu} j_{\nu} m_{\nu} \vert \bm{\alpha} \cdot \bm{A}^{\text{tw}} (\bm{r} + \bm{b}) \vert n_i j_i m_i \rangle}{E_i - E_{\nu} + \omega} \notag \\
	& + \sum_{n_{\nu} j_{\nu} m_{\nu}} \frac{\langle n_f j_f m_f \vert \bm{\alpha} \cdot \bm{A}^{\text{tw}} (\bm{r} + \bm{b}) \vert n_{\nu} j_{\nu} m_{\nu} \rangle   \langle n_{\nu} j_{\nu} m_{\nu} \vert \bm{\alpha} \cdot \bm{A}^{\text{pl} *} (\bm{r}) \vert n_i j_i m_i \rangle}{E_i - E_{\nu} - \omega}  \, ,
\end{align}
\end{widetext}
where $\vert n_i j_i m_i \rangle$ and $\vert n_f j_f m_f \rangle$ denote the states of the hydrogenlike ion before and after the scattering, and where $j_{i,f}$ and $m_{i,f}$ refer to the total angular momenta and their projections, and $n_{i,f}$ stand for principal quantum numbers. We here restrict ourselves to the nonresonant elastic scattering of the photons with the energy $\omega$ on the ground state of atoms. This implies that the total energy of the bound electron for the initial and final states of the atom with $n_i = n_f$ and $j_i = j_f$ obeys the energy conservation law $E_i = E_f$, and that the photon energy $\omega$ is not close to possible excitations of any intermediate states $\vert n_{\nu} j_{\nu} m_{\nu} \rangle$ over which the summation in the matrix element \eqref{eq:matrix1} is carried out, i.e., $\omega \neq E_{\nu} - E_i$.

In the matrix element \eqref{eq:matrix1}, the interaction of the atomic electrons with an incident Bessel beam is described by the transition operator $\bm{\alpha} \cdot \bm{A}^{\text{tw}} (\bm{r} + \bm{b})$, where $\bm{\alpha}$ denotes the vector of the Dirac matrices and $\bm{A}^{\text{tw}}$ is the vector potential of the beam as given by Eq. \eqref{eq:vec1}. Here, the impact parameter $\bm{b} $ occurs because the electron (coordinates) is shifted with regard to the beam axis. Since equation \eqref{eq:vec2} implies that the Bessel beam exhibits an inhomogeneous intensity distribution and a ringlike pattern in the transverse plane (cf. Fig. \ref{fig:geometry}), the Rayleigh scattering will explicitly depend on the atomic impact parameter $\bm{b} = (b_x,  b_y,  0)$. 

In Eq. \eqref{eq:matrix1}, we assumed that the scattered photons are plane waves $\bm{A}^{\text{pl}} (\bm{r}) = \bm{e}_{\bm{k}_2 \lambda_2} \, e^{i \bm{k}_2 \bm{r}}$ with $k_2 = k_1 = \omega / c$  measured by a detector placed at asymptotic distance under the direction $\bm{k}_2$. This is a reasonable assumption since all presently available detectors are plane wave detectors. To further analyze the transition amplitude \eqref{eq:matrix1}, we can decompose the plane-wave components of the incident and outgoing radiation in terms of the electric and magnetic multipole fields. When the wave vectors $\hat{\bm{k}}_1 = (\theta_{k_1}, \phi_{k_1})$ and $\hat{\bm{k}}_2 = (\theta_{k_2}, \phi_{k_2})$ are not both directed along the quantization $z$ axis, this decomposition may be written as
\begin{align}
\label{eq:mult}
	\bm{e}_{\bm{k} \lambda} \, e^{i \bm{k} \bm{r} } =& \sqrt{2 \pi} \sum_{L M} \sum_{p=0,1} i^L \, \sqrt{2L+1} \notag \\ 
	\times &(i \lambda )^{p} \, D_{M \lambda}^L (\phi_k, \theta_k, 0) \, \bm{a}_{L M}^{p} (\bm{r}) \, .
\end{align}
Here $D_{M \lambda}^L$ is the Wigner $D$ function, and $\bm{a}_{L M}^{p} (\bm{r})$ refers to the magnetic ($p=0$) and electric ($p=1$) multipole components, respectively \cite{Rose:1957}. If we substitute the multipole expansion \eqref{eq:mult} into Eq. \eqref{eq:matrix1} and make use of the vector potential \eqref{eq:vec1} of Bessel beams, we can rewrite the transition amplitude as
\begin{widetext}
\begin{align}
\label{eq:matrix2}
	\mathcal{M}_{m_f m_i}^{\lambda_2 \lambda_1} (\bm{b}) =& \sum_{M_1} \int a_{\varkappa m} (\bm{k}_{\perp_1}) \, e^{-i M_1 \phi_{k_1} + i \bm{k}_{\perp_1} \bm{b}} \, T_{m_f m_i}^{\lambda_2 \lambda_1} (M_1)  \, \frac{d^2 \bm{k}_{\perp_1}}{(2\pi)^2}  \, 
\end{align}
with the function $ T_{m_f m_i}^{\lambda_2 \lambda_1} (M_1)$ of the form
\begin{align}
\label{eq:matrix3}
	T_{m_f m_i}^{\lambda_2 \lambda_1} (M_1) = \sum_{L_1 p_1} &\sum_{L_2 M_2 p_2} 2 \pi i^{L_1 - L_2} \, \sqrt{(2L_1 + 1)(2L_2 + 1)} \, (i \lambda_1)^{p_1} (-i \lambda_2)^{p_2} \, e^{i M_2 \phi_{k_2}} \, d^{L_1}_{M_1 \lambda_1} (\theta_{k_1}) d^{L_2}_{M_2 \lambda_2} (\theta_{k_2}) \notag \\
	& \times \sum_{j_\nu} \left( \frac{\langle j_i m_i, L_1 M_1 \vert  j_{\nu} m_{\nu} \rangle   \langle j_{\nu} m_{\nu}, L_2 M_2 \vert j_f m_f \rangle}{\sqrt{(2 j_{\nu} + 1)(2 j_f + 1)}} \, S^{j_{\nu}}_{L_2 p_2, L_1 p_1} (\omega) \right. \notag \\
	& \left. \;\;\;\;\;\;\;\;\;\;\; + \frac{\langle j_i m_i,  L_2 M_2 \vert  j_{\nu} m_{\nu} \rangle   \langle j_{\nu} m_{\nu}, L_1 M_1 \vert j_f m_f \rangle}{\sqrt{(2 j_{\nu} + 1)(2 j_f + 1)}} \, S^{j_{\nu}}_{L_1 p_1, L_2 p_2} (- \omega) \right) \, ,
\end{align}
where we have used the Wigner small $d$ function and the Wigner-Eckart theorem \cite{Varshalovich:1988}. The reduced second-order matrix element is given by
\begin{align}
\label{eq:red}
	&S^{j_{\nu}}_{L_1 p_1, L_2 p_2} (\pm \omega) = \sum_{n_{\nu}} \frac{\langle n_f j_f \Vert \bm{\alpha} \cdot \bm{a}_{L_1}^{p_1} \Vert n_{\nu} j_{\nu} \rangle   \langle n_{\nu} j_{\nu} \Vert \bm{\alpha} \cdot \bm{a}_{L_2}^{p_2} \Vert n_i j_i \rangle}{E_i - E_{\nu} \pm \omega} \, .
\end{align}
\end{widetext}

To further simplify the matrix element \eqref{eq:matrix2}, we perform the integration over $k_{\perp_1}$ and $\phi_{k_1}$ with the help of Eq. \eqref{eq:amp} and by making use of the integral representation of the Bessel function \cite{Peshkov/PRA:2017}
\begin{align}
\label{eq:BesInt}
    &\frac{1}{2\pi} \int_0^{2\pi} e^{i (m - M_1) \phi_{k_1} + i \varkappa b \cos (\phi_{k_1} - \phi_b) } \, d \phi_{k_1} \notag \\
    &= i^{m - M_1} \, e^{i (m-M_1) \phi_b} \, J_{m-M_1} (\varkappa b) \, .
\end{align}
With this substitution, the transition amplitude for the scattering on a single hydrogenlike ion can be written as
\begin{align}
\label{eq:matrix4}
	\mathcal{M}_{m_f m_i}^{\lambda_2 \lambda_1} (\bm{b}) =& \sqrt{\frac{\varkappa}{2\pi}} \sum_{M_1} (-i)^{M_1} \, e^{i (m - M_1) \phi_b} \notag \\
	& \times  J_{m-M_1} (\varkappa b)  \, T_{m_f m_i}^{\lambda_2 \lambda_1} (M_1) \, .
\end{align}
As seen from this formula, the amplitude for the scattering of a Bessel beam depends not only on its helicity $\lambda_1$, the opening angle $\theta_{k_1}$ and the projection $m$ of the TAM, but also on the impact parameter $\bm{b}$ of the atom with respect to the beam axis. Below, we shall apply this transition amplitude to calculate the polarization of scattered light.

\subsection{\label{sec:single} Scattering on a single atom}
To characterize the polarization of scattered photons, we need to introduce the photon density matrix. For the scattering of twisted light on a single initially unpolarized atom with the impact parameter $\bm{b}$, the density matrix of scattered photons can be expressed in terms of the transition amplitudes as \cite{Balashov:2000}
\begin{align}
\label{eq:density1s}
	\langle \bm{k}_2 \lambda_2 \vert \hat{\rho}_{\gamma_2} \vert \bm{k}_2 \lambda'_2 \rangle &= \frac{1}{2 j_i + 1} \sum_{\lambda_1 \lambda'_1} \sum_{m_i m_f} \mathcal{M}_{m_f m_i}^{\lambda_2 \lambda_1} (\bm{b}) \notag \\
	&  \times   \mathcal{M}_{m_f m_i}^{\lambda'_2 \lambda'_1 \, *} (\bm{b}) \, \langle \bm{k}_1 \lambda_1 \vert \hat{\rho}_{\gamma_1} \vert \bm{k}_1 \lambda'_1 \rangle \, .
\end{align}
Here we assume that the magnetic sublevel population of the final state $\vert n_f j_f \rangle$ of the atom remains unobserved. The density matrix of an incident photon is $\langle \bm{k}_1 \lambda_1 \vert \hat{\rho}_{\gamma_1} \vert \bm{k}_1 \lambda'_1 \rangle = \delta_{\lambda_1 \lambda'_1}$ for a completely polarized radiation with the helicity $\lambda_1$. In typical experiments, however, the incident light is often unpolarized, i.e., the beam consists out of a mixture of photons in states of opposite helicity $\lambda_1 = \pm 1$ with equal intensities whose density matrix is $\langle \bm{k}_1 \lambda_1 \vert \hat{\rho}_{\gamma_1} \vert \bm{k}_1 \lambda'_1 \rangle = 1/2 \, \delta_{\lambda_1 \lambda'_1} \delta_{\lambda_1 +1} + 1/2 \, \delta_{\lambda_1 \lambda'_1} \delta_{\lambda_1 -1}$. Using the explicit expression of the amplitude \eqref{eq:matrix4}, we can rewrite the density matrix of scattered photons in the form
\begin{widetext}
\begin{align}
\label{eq:density2s}
	\langle \bm{k}_2 \lambda_2 \vert \hat{\rho}_{\gamma_2} \vert \bm{k}_2 \lambda'_2 \rangle  =& \frac{1}{2 j_i + 1} \frac{\varkappa}{2\pi} \sum_{\lambda_1 \lambda'_1} \sum_{m_i m_f} \sum_{M_1 M'_1} i^{M'_1 - M_1} \, e^{i (M'_1 - M_1) \phi_b}  \notag \\
	& \times T_{m_f m_i}^{\lambda_2 \lambda_1} (M_1) \, T_{m_f m_i}^{\lambda'_2 \lambda'_1 \, * } \, (M'_1) \, J_{m-M_1} (\varkappa b) \, J_{m-M'_1} (\varkappa b) \, \langle \bm{k}_1 \lambda_1 \vert \hat{\rho}_{\gamma_1} \vert \bm{k}_1 \lambda'_1 \rangle \, .
\end{align}

Let us analyze the special case of atoms placed right on the beam axis ($b=0$). In this scenario, the Bessel function from Eq. \eqref{eq:density2s} is just $J_{m-M_1} (0) = \delta_{m M_1}$, so that the photon density matrix reads
\begin{align}
\label{eq:density3s}
	\langle \bm{k}_2 \lambda_2 \vert \hat{\rho}_{\gamma_2} \vert \bm{k}_2 \lambda'_2 \rangle = \frac{1}{2 j_i + 1} \frac{\varkappa}{2\pi} \sum_{\lambda_1 \lambda'_1} \sum_{m_i m_f}  T_{m_f m_i}^{\lambda_2 \lambda_1} (M_1 = m) \, T_{m_f m_i}^{\lambda'_2 \lambda'_1 \, * } \, (M'_1 = m) \, \langle \bm{k}_1 \lambda_1 \vert \hat{\rho}_{\gamma_1} \vert \bm{k}_1 \lambda'_1 \rangle  \, .
\end{align}
\end{widetext}
This expression indicates that the atom on the beam axis can just absorb a photon with the projection of the angular momentum $m$ \cite{Schmiegelow/NC:2016, Peshkov/PS:2016}. In practice, however, it is difficult to position the atom just \textit{on} the beam axis ($b=0$). Therefore, in the next section we will consider the scattering of twisted light by a mesoscopic atomic target in which atoms are localized with nanometer precision. 

\subsection{\label{sec:mesoscopic} Scattering on a mesoscopic atomic target}
\begin{figure}[t!]
\includegraphics[width=0.89\linewidth]{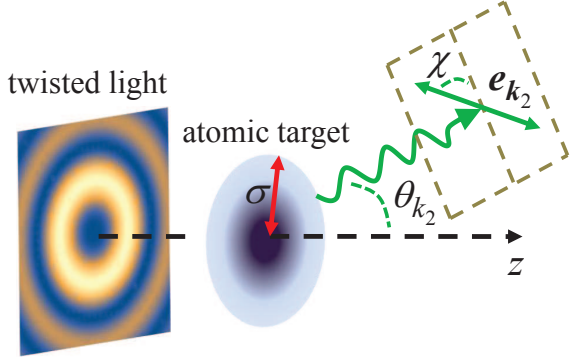}
\caption{Geometry of the Rayleigh scattering of twisted light by a mesoscopic atomic target of size $\sigma$. While the quantization $(z)$ axis is taken along the propagation direction of the incident beam, the center of atomic target is placed on the beam axis. The emission direction of the outgoing photons is characterized by the angle $\theta_{k_2}$, and their polarization vector $\bm{e}_{\bm{k}_2}$ is described by the angle $\chi$.}
\label{fig:geometry}
\end{figure}

The experiments on the interaction of twisted light beams with the atoms or ions, which are localized in a small volume of several tens of nanometers by means of a microstructured Paul trap, are feasible today \cite{Schmiegelow/NC:2016}. For the Rayleigh scattering by such a mesoscopic atomic target centered on the beam axis, the density matrix of scattered photons is given by \cite{Peshkov/PS:2016, Surzhykov/PRA:2016}
\begin{align}
\label{eq:density1m}
	&\langle \bm{k}_2 \lambda_2 \vert \hat{\rho}_{\gamma_2} \vert \bm{k}_2 \lambda'_2 \rangle = \frac{1}{2 j_i + 1} \sum_{\lambda_1 \lambda'_1} \sum_{m_i m_f} \langle \bm{k}_1 \lambda_1 \vert \hat{\rho}_{\gamma_1} \vert \bm{k}_1 \lambda'_1 \rangle \notag \\
	&\;\;\;\;  \times \int f(\bm{b}) \mathcal{M}_{m_f m_i}^{\lambda_2 \lambda_1} (\bm{b}) \, \mathcal{M}_{m_f m_i}^{\lambda'_2 \lambda'_1 \, *} (\bm{b}) \, d^2 \bm{b} \, ,
\end{align}
where the atomic density of this target in the transverse plane (cf. Fig. \ref{fig:geometry}) is assumed to follow the Gaussian distribution \cite{Eschner/EPJD:2003}
\begin{align}
\label{eq:distr}
	f(\bm{b}) = \frac{1}{2\pi \sigma^2} e^{- \frac{\bm{b}^2}{2 \sigma^2}} \, .
\end{align}
In this formula, $\sigma$ is the width of the target. After making use of the transition amplitude \eqref{eq:matrix4} and integrating over the azimuthal angle $\phi_b$, the photon density matrix for the mesoscopic atomic target becomes
\begin{align}
\label{eq:density2m}
	&\langle \bm{k}_2 \lambda_2 \vert \hat{\rho}_{\gamma_2} \vert \bm{k}_2 \lambda'_2 \rangle = \frac{1}{2 j_i + 1} \frac{\varkappa}{2\pi \sigma^2} \notag \\
	& \;\; \times \sum_{\lambda_1 \lambda'_1} \sum_{m_i m_f M_1}  \langle \bm{k}_1 \lambda_1 \vert \hat{\rho}_{\gamma_1} \vert \bm{k}_1 \lambda'_1 \rangle \,  T_{m_f m_i}^{\lambda_2 \lambda_1} (M_1) \notag \\
	& \;\; \times T_{m_f m_i}^{\lambda'_2 \lambda'_1 \, * } (M_1) \int_{0}^{\infty} J_{m-M_1}^2 (\varkappa b) \, e^{- \frac{b^2}{2 \sigma^2}} \, b db \, .
\end{align}

Both the density matrices \eqref{eq:density3s} and \eqref{eq:density2m} show that the polarization of outgoing photons depends on the TAM projection $m$ of an incident Bessel beam in the elastic scattering by a single atom or by a mesoscopic atomic target. However, there is no $m$ dependence for a rather large macroscopic atomic target, as we shall see below.  
\subsection{\label{sec:macroscopic} Scattering on a macroscopic atomic target}
We next analyze the scattering of Bessel beam by a macroscopic target in which atoms are distributed randomly over the whole extent of the incident beam. In the case of such a large target, the photon density matrix is defined by \cite{Surzhykov/PRA:2015}
\begin{align}
\label{eq:density1Ma}
	&\langle \bm{k}_2 \lambda_2 \vert \hat{\rho}_{\gamma_2} \vert \bm{k}_2 \lambda'_2 \rangle = \frac{1}{2 j_i + 1} \sum_{\lambda_1 \lambda'_1} \sum_{m_i m_f} \langle \bm{k}_1 \lambda_1 \vert \hat{\rho}_{\gamma_1} \vert \bm{k}_1 \lambda'_1 \rangle \notag \\
	& \; \times \int \mathcal{M}_{m_f m_i}^{\lambda_2 \lambda_1} (\bm{b}) \, \mathcal{M}_{m_f m_i}^{\lambda'_2 \lambda'_1 \, *} (\bm{b}) \, d^2 \bm{b} \notag \\
	& =  \frac{1}{2 j_i + 1} \sum_{\lambda_1 \lambda'_1}  \sum_{m_i m_f} \sum_{M_1 M'_1} \langle \bm{k}_1 \lambda_1 \vert \hat{\rho}_{\gamma_1} \vert \bm{k}_1 \lambda'_1 \rangle \notag \\
	& \; \times \int a_{\varkappa m} (\bm{k}_{\perp_1}) \, a_{\varkappa m}^{*} (\bm{k}'_{\perp_1}) \, e^{-i M_1 \phi_{k_1} + i M'_1 \phi_{k'_1} + i ( \bm{k}_{\perp_1} - \bm{k}'_{\perp_1} ) \bm{b}} \notag \\
	& \; \times  T_{m_f m_i}^{\lambda_2 \lambda_1} (M_1) \, T_{m_f m_i}^{\lambda'_2 \lambda'_1 \, *} (M'_1)  \, \frac{d^2 \bm{k}_{\perp_1} d^2 \bm{k}'_{\perp_1} d^2 \bm{b}}{(2\pi)^4} \, ,
\end{align}
where we have used the transition amplitude \eqref{eq:matrix2}. Here the integration over the impact parameter $\bm{b}$ yields immediately the $\delta$ function $\delta (\bm{k}_{\perp_1} - \bm{k}'_{\perp_1})$. Moreover, if we perform the integration over the wave vector $\bm{k}'_{\perp_1}$ and over the azimuthal angle $\phi_{k_1}$, we simply obtain $M_1 = M'_1$. We can further simplify the photon density matrix \eqref{eq:density1Ma} by integrating over $k_{\perp_1}$, so that
\begin{align}
\label{eq:density2Ma}
	\langle \bm{k}_2 \lambda_2 \vert \hat{\rho}_{\gamma_2} \vert \bm{k}_2 \lambda'_2 \rangle &= \frac{1}{2 j_i + 1} \sum_{\lambda_1 \lambda'_1} \sum_{m_i m_f M_1} T_{m_f m_i}^{\lambda_2 \lambda_1} (M_1) \notag \\
	&  \times  T_{m_f m_i}^{\lambda'_2 \lambda'_1 \, * } (M_1) \, \langle \bm{k}_1 \lambda_1 \vert \hat{\rho}_{\gamma_1} \vert \bm{k}_1 \lambda'_1 \rangle \, .
\end{align}
This formula shows that in the scattering on a macroscopic target the density matrix of outgoing photons and, hence, also their polarization are independent of the TAM projection $m$ of incoming twisted light, but still depend on its helicity $\lambda_1$ and opening angle $\theta_{k_1}$.

\subsection{\label{sec:polarization} Polarization parameters}
With the photon density matrices obtained above, we can now analyze the polarization of the Rayleigh scattered light. As usual in atomic and optical physics, the polarization properties of photons are characterized by the Stokes parameters \cite{Balashov:2000}. In particular, the parameter $P_1 = (I_{\chi = 0^{\circ}} - I_{\chi = 90^{\circ}}) / (I_{\chi = 0^{\circ}} + I_{\chi = 90^{\circ}})$ characterizes the degree of linear polarization and is determined by the intensities $I_{\chi}$ of scattered light linearly polarized at an angle $\chi = 0^{\circ}$ or $\chi = 90^{\circ}$. Here the angle $\chi$ is defined with respect to the plane spanned by the directions of incident and outgoing photons (cf. Fig. \ref{fig:geometry}). Another parameter $P_2$, given by a similar ratio but for $\chi = 45^{\circ}$ and $\chi = 135^{\circ}$, is close to zero and therefore is not of interest. On the other hand, the nonzero parameter $P_3 = (I_{\lambda_2 = +1} - I_{\lambda_2 = -1}) / (I_{\lambda_2 = +1} + I_{\lambda_2 = -1})$ characterizes the degree of circular polarization and is determined by the intensities $I_{\lambda_2}$ of outgoing circularly polarized photons with the helicity $\lambda_2 = \pm 1$. Both these Stokes parameters can be expressed in terms of the density matrix of photons as \cite{Balashov:2000}
\begin{widetext}
\begin{align}
\label{eq:stokes1}
	&P_1 (\theta_{k_2}) = - \frac{ \langle \bm{k}_2 \lambda_2 = +1 \vert \hat{\rho}_{\gamma_2} \vert \bm{k}_2 \lambda'_2 = -1 \rangle + \langle \bm{k}_2 \lambda_2 = -1 \vert \hat{\rho}_{\gamma_2} \vert \bm{k}_2 \lambda'_2 = +1 \rangle}{\langle \bm{k}_2 \lambda_2 = +1 \vert \hat{\rho}_{\gamma_2} \vert \bm{k}_2 \lambda'_2 = +1 \rangle + \langle \bm{k}_2 \lambda_2 = -1 \vert \hat{\rho}_{\gamma_2} \vert \bm{k}_2 \lambda'_2 = -1 \rangle} \, , 
\end{align}
\begin{align}
\label{eq:stokes3}	
	&P_3 (\theta_{k_2}) = \frac{ \langle \bm{k}_2 \lambda_2 = +1 \vert \hat{\rho}_{\gamma_2} \vert \bm{k}_2 \lambda'_2 = +1 \rangle - \langle \bm{k}_2 \lambda_2 = -1 \vert \hat{\rho}_{\gamma_2} \vert \bm{k}_2 \lambda'_2 = -1 \rangle}{\langle \bm{k}_2 \lambda_2 = +1 \vert \hat{\rho}_{\gamma_2} \vert \bm{k}_2 \lambda'_2 = +1 \rangle + \langle \bm{k}_2 \lambda_2 = -1 \vert \hat{\rho}_{\gamma_2} \vert \bm{k}_2 \lambda'_2 = -1 \rangle} \, .
\end{align}
\end{widetext}
As seen from these expressions, the Stokes parameters depend on the direction $\theta_{k_2}$ of scattered light. Therefore, in Sec. \ref{sec:results} we will use Eqs. \eqref{eq:stokes1} and \eqref{eq:stokes3} to investigate the polarization of outgoing photons for different scattering angles $\theta_{k_2}$.
\begin{figure*}[t!]
\includegraphics[width=0.89\linewidth]{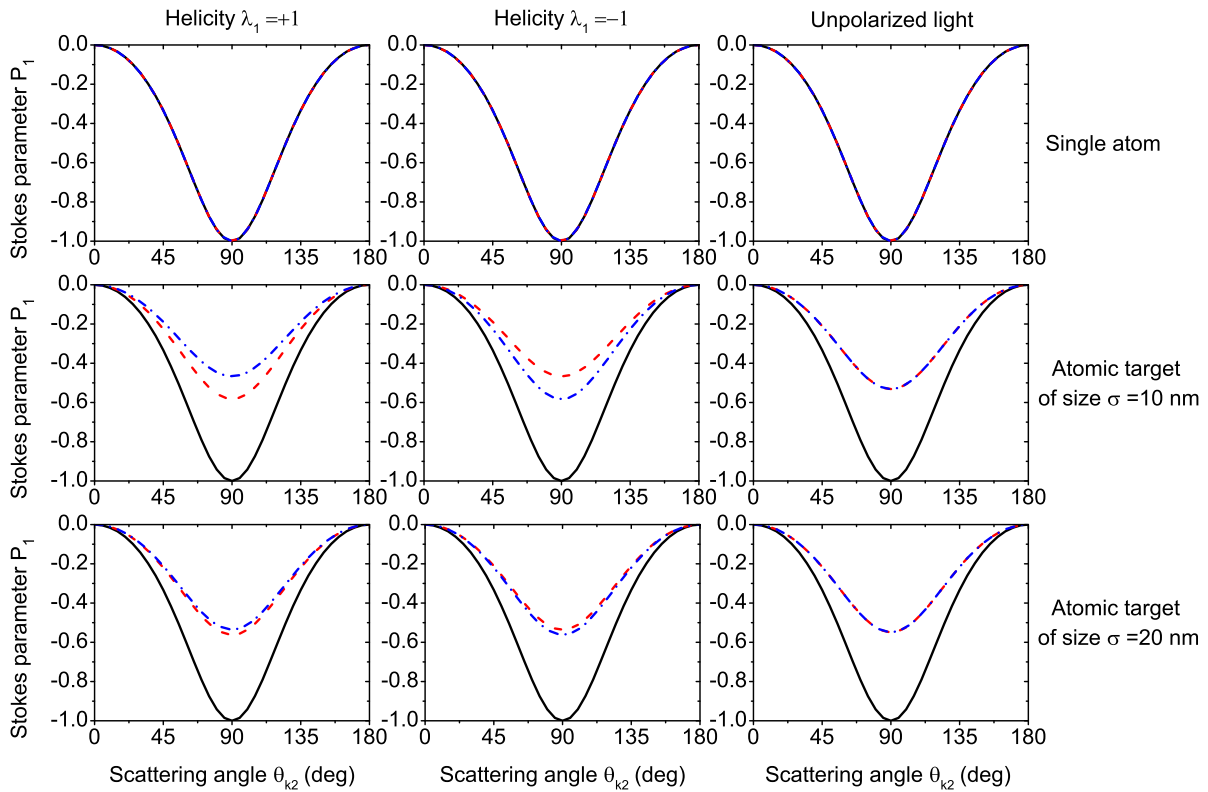}
\caption{Stokes parameters $P_1$ of Rayleigh scattered light on hydrogenlike C$^{5+}$ ions in their ground state as a function of the emission angle $\theta_{k_2}$. Results for incident plane waves (black solid lines) are compared with those for Bessel beams with TAM $m=+1$ (red dashed lines) and $m=-1$ (blue dash-dotted lines), respectively. Relativistic calculations were performed for a single atom (top row) and for mesoscopic atomic targets of size $\sigma = 10$ nm (middle row) and $\sigma = 20$ nm (bottom row), which are centered on the beam axis. Results are shown for different helicities $\lambda_1$ of the incident light: $\lambda_1 = +1$ (left column), $\lambda_1 = -1$ (central column), and for the unpolarized light (right column). Both the opening angle $\theta_{k_1} = 30^{\circ}$ of Bessel beams and the photon energy $\hbar \omega = 100$ eV are kept fixed.}
\label{fig:p1}
\end{figure*}
\begin{figure*}[t!]
\includegraphics[width=0.89\linewidth]{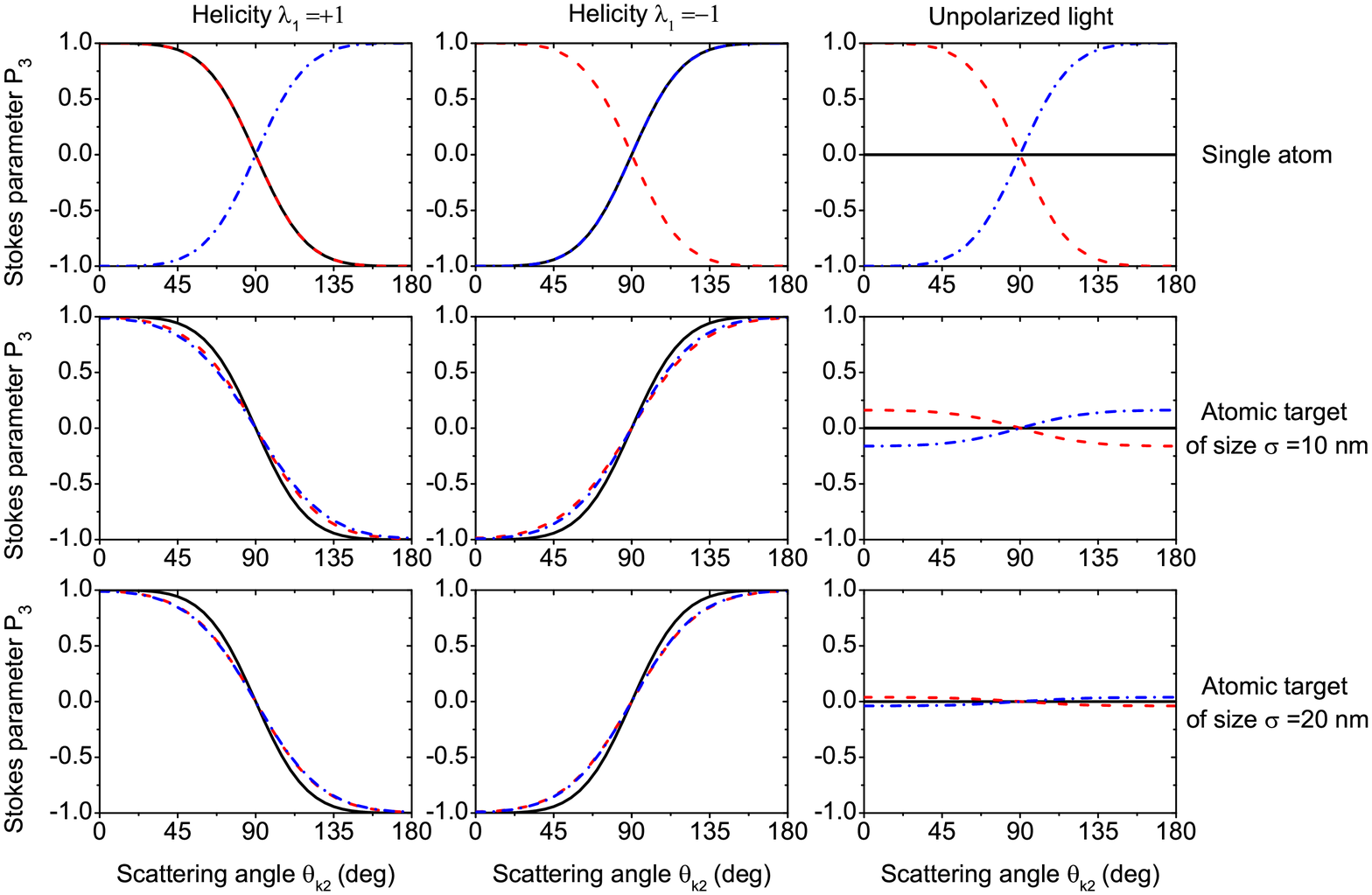}
\caption{Same as Fig. \ref{fig:p1}, but for the Stokes parameters $P_3$ of elastically scattered photons.} 
\label{fig:p3}
\end{figure*}
\subsection{\label{sec:computations} Computations}
Before we present our results for the Stokes parameters, let us briefly discuss some computational details. The evaluation of the polarization of scattered photons requires the knowledge of the reduced second-order transition amplitude \eqref{eq:red}, which involves the summation over the complite basis of the intermediate states $\vert n_{\nu} j_{\nu} \rangle$. In order to perform this summation, we use two independent approaches: the finite basis-set method and the Dirac-Coulomb Green's function (see Ref. \cite{Volotka/PRA:2016} for further details). These two numerical methods provide identical results, which demonstrates the high accuracy of our calculations.
\section{\label{sec:results}Results and discussion}
In the previous sections we found the Stokes parameters $P_1$ and $P_3$ describing the polarization of scattered photons in the Rayleigh scattering of twisted Bessel beams by hydrogenlike ions. Such polarization parameters can be observed in current experiments \cite{Blumenhagen/NJP:2016} and are expressed in terms of the photon density matrix, as seen from Eqs. \eqref{eq:stokes1} and \eqref{eq:stokes3}. We further analyze how these Stokes parameters of scattered light depend on its emission angle $\theta_{k_2}$ for incident Bessel beams with different projections $m$ of the TAM, helicities $\lambda_1$, and opening angles $\theta_{k_1}$. In addition, we compare these parameters $P_1$ and $P_3$ for twisted light with those obtained for a plane-wave radiation of the same helicity incident along the $z$ axis. Calculations were performed for the photon energy $\hbar \omega = 100$ eV and for three different targets of C$^{5+}$ ions: a single atom \eqref{eq:density3s}, a mesoscopic target \eqref{eq:density2m}, and a macroscopic target \eqref{eq:density2Ma} that are centered on the beam axis.
\subsection{\label{sec:pol sing mes} Polarization for a single atom and mesoscopic atomic target}
We start with the first Stokes parameter $P_1$ that characterizes the degree of linear polarization of outgoing photons. Fig. \ref{fig:p1} illustrates the parameter $P_1$ as a function of the emission angle $\theta_{k_2}$ for the Rayleigh scattering on a single atom (top row) as well as on the mesoscopic targets of size $\sigma = 10$ nm (middle row) and $\sigma = 20$ nm (bottom row). As seen from this figure, the outgoing photons are completely $P_1 = -1$ linearly polarized in the $\chi = 90^{\circ}$ direction at the scattering angle $\theta_{k_2} = 90^{\circ}$ for incoming plane waves (black solid lines). This is also true if a Bessel beam collides with a single atom located on the beam axis. However, the scattering of such a Bessel beam by mesoscopic target with width $\sigma = 10$ nm, for example, leads to a significant decrease of the polarization at the angle $\theta_{k_2} = 90^{\circ}$, namely $P_1 = -0.58$ when $m=+1$ (red dashed line) or $P_1 = -0.47$ when $m=-1$ (blue dash-dotted line) for positive helicity $\lambda_1 = +1$, and vice versa for negative helicity $\lambda_1 = -1$. Thus the Stokes parameter $P_1$ of scattered photons depends on the TAM projection $m$ of twisted light of a well-defined helicity $\lambda_1$ in the scattering by a mesoscopic target. On the other hand, $P_1$ is independent of TAM $m$ if an incoming Bessel beam is unpolarized (cf. Fig. \ref{fig:p1}).

\begin{figure*}[t!]
\includegraphics[width=0.79\linewidth]{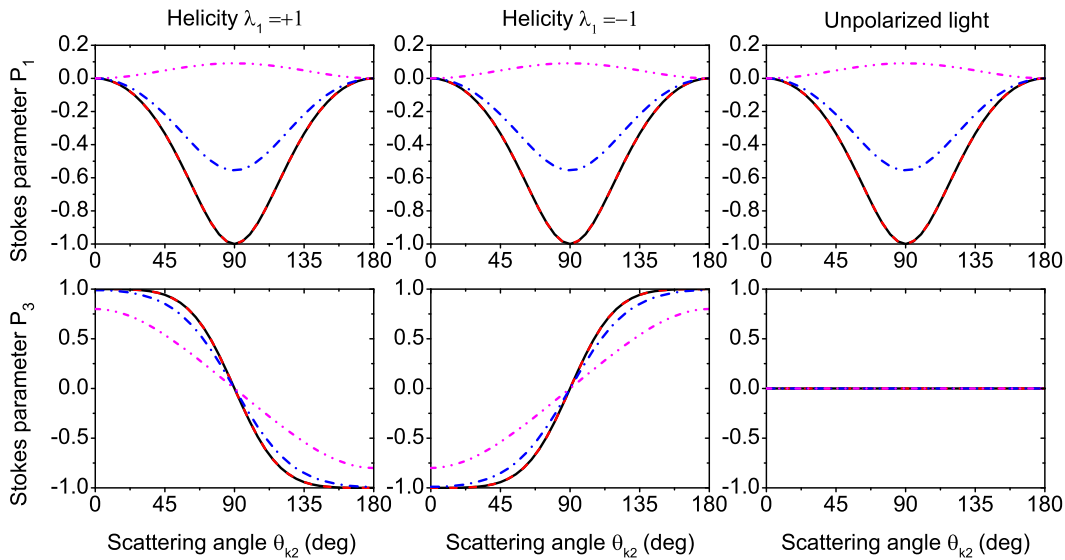}
\caption{Stokes parameters $P_1$ (top row) and $P_3$ (bottom row) of elastically scattered photons on hydrogenlike C$^{5+}$ ions in their ground state for a macroscopic target. Plane-wave results (black solid lines) are compared with those for Bessel beams with opening angles $\theta_{k_1} = 1^{\circ}$ (red dashed lines), $\theta_{k_1} = 30^{\circ}$ (blue dash-dotted lines), and $\theta_{k_1} = 60^{\circ}$ (magenta dash-dot-dotted lines). Calculations were performed for different helicities $\lambda_1$ of the incident light: $\lambda_1 = +1$ (left column), $\lambda_1 = -1$ (central column), and for the unpolarized light (right column), when the photon energy $\hbar \omega = 100$ eV is fixed.}
\label{fig:macr}
\end{figure*}

Up to this point, we have discussed the linear polarization of elastically scattered light. In order to analyze its degree of circular polarization, the third Stokes parameter $P_3$ as a function of the scattering angle $\theta_{k_2}$ is presented in Fig. \ref{fig:p3}. One sees that when the incident radiation is a plane wave of helicity $\lambda_1$, the photons scattered in the forward ($\theta_{k_2} = 0^{\circ}$) direction are completely circularly polarized, namely $P_3 = 1$ if $\lambda_1 = 1$ or $P_3 = -1$ if $\lambda_1 = -1$. Moreover, the Stokes parameter $P_3$ of outgoing photons for the scattering of a twisted beam by a single atom on the beam axis coincides with the plane-wave results at all emission angles $\theta_{k_2}$ if the TAM projection of the beam is $m = \lambda_1$, as shown in Fig. \ref{fig:p3}. However, $P_3$ corresponding to twisted light shows the opposite behaviour to $P_3$ for the plane waves if the TAM projection is $m = - \lambda_1$. Such a difference in the polarization (or helicity) of outgoing photons is caused by the conservation of the angular momentum projection: the helicity $\lambda_2$ of a photon emitted in the forward ($\theta_{k_2} = 0^{\circ}$) direction should be equal to the projection $M_1$ of the angular momentum of a photon absorbed by the atom on the beam axis, which is $M_1 = m$ for a Bessel beam \eqref{eq:density3s}, in contrast to $M_1 = \lambda_1$ for a plane wave.

Let us consider how the mesoscopic atomic target may affect the third Stokes parameter of scattered light. Eq. \eqref{eq:density2m} implies that all possible projections $M_1$ of the angular momentum of incoming photons are able to contribute to the scattering of twisted light by mesoscopic target, in contrast to $M_1 = m$ for the scattering by a single atom. As a result, in the case of a mesoscopic target the parameter $P_3$ of outgoing photons for an incident Bessel beam is slightly different from that for a plane wave in the angular range $30^{\circ} \lesssim \theta_{k_2} \lesssim 70^{\circ}$ and  $110^{\circ} \lesssim \theta_{k_2} \lesssim 150^{\circ}$, as can be seen from the middle and bottom rows of Fig. \ref{fig:p3}. In addition, the Stokes parameters $P_1$ and $P_3$ are quite different for the two TAM projections $m = \pm 1$ of the beam when the mesoscopic target is rather small ($\sigma = 10$ nm). However, Figs. \ref{fig:p1} and \ref{fig:p3} also show that this difference between the Stokes parameters for various TAM $m$ decreases with increasing size of the target ($\sigma = 20$ nm).

Strong effects of ``twistedness'' in the polarization of scattered light can be observed also for an incoming unpolarized Bessel beam containing photons of both helicities $\lambda_{1} = \pm 1$ but with a fixed TAM projection $m$. In particular, Fig. \ref{fig:p3} demonstrates that the Stokes parameter $P_3$ of outgoing photons is not always zero in the scattering of such a beam, in contrast to $P_3$ for incident unpolarized plane waves. For example, when the unpolarized twisted light with TAM projection $m= +1$ collides with a single atom, the third Stokes parameter (red dashed line) behaves similarly to that obtained for the incident beam with a well-defined helicity $\lambda_1 = + 1$. This is because in the scattering of twisted light by a single atom $P_3$ does not depend on the helicity $\lambda_{1}$, but is only sensitive to the TAM $m$. With increasing target size $\sigma$, however, the parameter $P_3$ for the case of unpolarized Bessel beam decreases and tends to zero as expected for incoming unpolarized plane waves (cf. Fig. \ref{fig:p3}).
\subsection{\label{sec:pol macr} Polarization for a macroscopic atomic target}
Finally, we consider the scattering of twisted light by a macroscopic target as it occurs, for instance, for the scattering at a foil \cite{Blumenhagen/NJP:2016}. For such an extended target, the polarization of outgoing photons is independent of the TAM projection $m$ of the twisted light, and as pointed out already in Sec. \ref{sec:macroscopic}. In Fig. \ref{fig:macr} we compare the two Stokes parameters $P_1$ and $P_3$ of the scattered light for different opening angles $\theta_{k_1}$ of Bessel beams with those for plane waves incident along the $z$ axis. Similar as before, results were obtained as a function of the scattering angle $\theta_{k_2}$ for different helicities of the radiation. Here one can see that the parameters $P_1$ and $P_3$ for the scattering of a Bessel beam with a very small opening angle ($\theta_{k_1} = 1^{\circ}$) are almost identical to those as obtained for an incident plane waves. However, the Stokes parameter $P_1$ behaves very differently for large opening angles ($\theta_{k_1} = 60^{\circ}$) and may become even positive at the emission angle $\theta_{k_2} = 90^{\circ}$. Moreover, for large angles $\theta_{k_1}$, the circular polarization of the scattered photons is decreased in forward direction, for example $P_3 = \pm 0.8$ if the helicity of a Bessel beam is $\lambda_1 = \pm 1$. These modifications of the polarization of scattered light follow from Eq. \eqref{eq:density2Ma} and imply that the scattering of a Bessel beam by macroscopic target can be considered as a scattering of plane waves propagating at the opening angle $\theta_{k_1}$ with respect to the quantization $z$ axis.
\section{\label{sec:summary}Summary and outlook}

In summary, we explore the Rayleigh scattering of twisted light by hydrogenlike ions within the framework of second-order perturbation theory and Dirac's relativistic equation. In this analysis, we focused on the polarization of photons scattered by a single atom, by a mesoscopic target (atoms in a trap), or by a macroscopic target (foil). The polarization Stokes parameters of outgoing photons were calculated especially for hydrogenlike carbon and for incident twisted Bessel beams. We have shown that the linear and circular polarization of scattered light depends generally on the helicity $\lambda_1$ and the opening angle $\theta_{k_1}$ of Bessel beams, leading to Stokes parameters that differs quite significantly from the scattering of incident plane-wave photons. Moreover, the polarization of the scattered photons is very sensitive to the TAM projection $m$ of twisted light for mesoscopic atomic targets of a few tens of nm in size, while it remains unaffected by the TAM $m$ in the case of a larger macroscopic target. Although our study was restricted to the scattering by hydrogenlike ions in their ground $1s$ state, similar polarization properties can also be observed in the scattering of twisted light by electrons in other $s$-shells. For example, we expect the same scattering polarization pattern for Ca$^{+}$ ions that were used in a recent experiment on the photoexcitation by twisted light \cite{Schmiegelow/NC:2016}. Thus the Rayleigh scattering may serve as an accurate technique for measuring the properties of twisted beams in a wide range of photon energies, and in particular at rather high energies.

The interaction of twisted light with atoms may lead not only to the scattering of photons, but also to the change in the atomic polarizability. The knowledge of the atomic polarizability induced by twisted radiation is very important in laser cooling and trapping experiments, and its analysis will be presented in a forthcoming publication.

\section*{\label{sec:acknowledgments}Acknowledgments}

This work was supported by the DFG priority programme ``Quantum Dynamics in Tailored Intense Fields''. A.A.P. acknowledges support from the Helmholtz Institute Jena and the Research School of Advanced Photon Science of Germany.

\vspace*{0.5cm}

\end{document}